\documentstyle[12pt,floats,epsf,epsfig,psfig,amssymb,aps,russian]{revtex}
\tighten

\newcommand{\Img}{\mathop{\rm Im}}
\newcommand{\Real}{\mathop{\rm Re}}
\newcommand{\reduction}[2]{#1 \biggr|_{#2}}
\begin{document}

\title{\uppercase{On the mechanism of formation of the Efimov states
in the helium $^4$H}\lowercase{e} \uppercase{trimer}}
\author{ E. A. Kolganova}
\address{ Laboratory of Computing Techniques and Automation\\
Joint Institute for Nuclear Research, 141980 Dubna,
Moscow region, Russia}
\author{ A. K. Motovilov\thanks{On leave of absence from
         the Laboratory of Theoretical Physics,
    Joint Institute  for Nuclear Research, Dubna, 141980, Russia}}
\address{Physikalishes Institut der Universit\"at Bonn \\
      Endenicher Allee 11\,--\,13, D\,--\,53115 Bonn, Germany}
\date{August 19, 1998} 
\maketitle
\bigskip

\begin{abstract}
A mechanism of disappearance and formation of the Efimov levels 
of the helium $^4$He trimer is studied when the force of the 
interatomic interaction is changed. It is shown that these 
levels arise from virtual levels which are in turn formed from 
(quasi)resonances settled on the real axis. The resonances 
including virtual levels are calculated by the method based on 
the solution of the boundary value problem, at complex energies, 
for the Faddeev differential equations describing the scattering 
processes $(2+1\to 2+1;\,1+1+1)$.  All the calculations are 
performed with the known interatomic Aziz He\,--\,He\,-\, 
potential HFD-B.  A very strong repulsive component of this 
potential at short distances between helium atoms is 
approximated by a hard core. A special attention is paid to the 
substantiation of the method used for computing resonances and 
to the investigation of its applicability range.
\end{abstract}
\vspace*{1cm}

\section{Introduction}
\label{Intro}

The $^4$He three-atomic system is of considerable interest in
various fields of physical chemistry and molecular physics.
Studies of the helium dimer and trimer represent an important
step towards understanding the properties of helium liquid
drops, superfluidity in  $^4$He films, and so on (see, for
instance, Refs.~\cite{RamaKrishna,LehmanScoles,GrebToeVil}).
Besides, the helium trimer is probably a unique system where
a direct manifestation of the Efimov effect ~\cite{VEfimov} can
be observed since the binding energy $\epsilon_d$ of the $^4$He
dimer is extremely small
($\sim-1$\,mK~\cite{DimerExp,DimerExp1,Science}) even in the
molecular scale.  For this reason, the helium trimer is
certainly of interest for nuclear physicists, too. Moreover a
theoretical study of the $^4$He trimer is based just on the same
methods of the theory of few-body systems that are used in
solving three--body nuclear problems.

From the standpoint of the general theory of few--body systems,
the $^4$He trimer belongs to three--body systems that are most
difficult for a specific investigation, first, owing to its
Efimov nature, and second, because it is necessary to take into
account the practically hard core in the interatomic
He\,--\,He\,-\,interaction ~\cite{Aziz79,Aziz87,Aziz91,Tang95}.
At the same time the problem of three helium atoms can be
considered as an example of an ideal three--body quantum problem
since the $^4$He atoms are identical neutral bosons with zero
spin and the analysis of this problem is complicated neither by
separation of spin--isospin variables nor by taking into account
the Coulomb interaction.

There is a great number of experimental and theoretical studies 
of $^4$He clusters. However, most of the theoretical 
investigations consist merely in computing the ground-state 
energies of clusters of that sort, mainly on the basis of 
variational methods \cite{V1,V2,V3,V4,V5}.   Besides, the 
methods based on hyperspherical expansions of the Schr\"odinger 
and Faddeev equations \cite{Levinger,EsryLinGreene,Nielsen} in 
the coordinate representation were used.  Also, the Faddeev 
integral equations in the momentum representation were employed 
in Refs.~\cite{Nakai,Gloeckle} while the results of 
Ref.~\cite{CGM} are based on a direct solving the 
two-dimensional Faddeev differential equations in configuration 
space. From the experimental studies we would like to mention 
those of Refs.  \cite{DimerExp,DimerExp1,Science,Toennis2} where 
clusters consisting of a small number of noble gase atoms were 
investigated.

Though much effort was undertaken for studying molecular
clusters various problems related to the $^4$He
trimer remained beyond the scope of thorough consideration. In
particular, the elastic scattering phases of a helium atom on a
helium dimer and breakup amplitudes (at ultralow
energies) have been calculated only
recently~\cite{KMS-PRA,MSK-CPL,KMS-JPB}.  These computations
were preceded only by the computation of characteristics of
the He--He$_2$ scattering at zero energy~\cite{Nakai} and estimation
of the recombination rate $(1+1+1\to2+1)$~\cite{Fed96}.

As a matter of fact, we have already pointed out basic reasons
for computations of excited states and scattering being
especially difficult in the $^4$He$_3$ system. First, this is a
low energy of the dimer $\epsilon_d$ which necessitates to
consider very large domains in the configuration space with a
characteristic size of hundreds of \AA. Second, a very
strong repulsive component in the He--He interaction
produces large errors in the standard approximation
of the three--atomic Hamiltonian at short distances between
atoms. The capacities of modern computers do not yet allow one to
reach dimensions of grids that would remove both the
above-mentioned reasons and would provide stable results with
the use of the conventional methods.

The present paper is a sequel of studies of the $^4$He$_3$
system undertaken in the papers~\cite{KMS-PRA,MSK-CPL,KMS-JPB}
within an approach that is capable, as we think, to resolve both
the above-mentioned numerical problems.  In these papers the
repulsive component of the He--He interaction at short distances
between atons is approximated by a hard core.  This allows one
to investigate the $^4$He$_3$ system within a mathematically
rigorous method of solving a three-body problem in the
Boundary-Condition Model developed in~\cite{MerMot,MMYa}. An
important advantage of such an approach that essentially
diminishes computational errors is the necessity to approximate,
inside the core domains, only the Laplacian operator instead of
the sum of this operator and a huge repulsive components of the
He\,--\,He\,-\,potentials (see \cite{KMS-JPB}).
In~\cite{KMS-PRA,MSK-CPL,KMS-JPB}, such an approach has been
successfully applied for calculating not only scattering but
also binding energies of the ground and excited states of the
helium trimer. Investigation made in
~\cite{KMS-PRA,MSK-CPL,KMS-JPB} has shown that the method
proposed in~\cite{MerMot,MMYa} is well suited for performing
three--body molecular computations in the case where repulsive
components of interatomic interactions are of a hard core
nature.

There is a series of works
\cite{EsryLinGreene,Gloeckle,KMS-JPB} showing
that the excited state of the $^4$He trimer is
initiated indeed by the Efimov effect ~\cite{VEfimov}.
In these works the various versions of
the Aziz $^4$He--$^4$He potential were employed
(HFDHE2~\cite{Aziz79}, HFD-B~\cite{Aziz87}, and
LM2M2~\cite{Aziz91}).  However, the basic result of
Refs.~\cite{EsryLinGreene,Gloeckle,KMS-JPB} on the excited state
of the helium trimer is the same: this state disappears when the
interatomic potential is multiplied by the ``amplification
factor" $\lambda$ of order 1.2.  More precisely, if this
potential is multiplied by the increasing factor $\lambda>1$ then the
following effect is observed. First, the difference
$\epsilon_d(\lambda)-E_t^{(1)}(\lambda)$ between the dimer
energy $\epsilon_d(\lambda)$  and the energy of the trimer excited
state $E_t^{(1)}(\lambda)$ increases. Then the behavior of this
difference radically changes and with further increase of
$\lambda$ it monotonously decreases. At $\lambda\approx 1.2$
the level $E_t^{(1)}$ disappears. It is just such a nonstandard
behavior of the energy $E_t^{(1)}(\lambda)$ as the coupling
between helium atoms becomes more and more strengthening,
points to the Efimov nature of the trimer excited state. And
vice versa, when $\lambda$ slightly decreases (no more than
2\,\%), the second excited state $E_t^{(2)}$ appears in the
trimer~\cite{EsryLinGreene,Gloeckle}.

This paper is aimed at elucidating the fate of the trimer
excited state upon its disappearance in the physical sheet when
$\lambda>1$ and at studying the mechanism of arising of new
excited states when $\lambda<1$. As the interatomic He\,--\,He
potential, we use the potential HFD-B~\cite{Aziz87}. We have
established that for such He\,--\,He\,-\,interactions the trimer
excited level $E_t^{(1)}$ merges with the threshold $\epsilon_d$
at $\lambda\approx1.18$ and with further decreasing $\lambda$ it
transforms into a virtual level of the first order (a simple
real pole of the analytic continuation of the scattering matrix)
lying in the unphysical energy sheet adjoining the physical
sheet along the spectral interval between $\epsilon_d$ and the
three--body threshold.  We trace the position of this level for
$\lambda$ increasing up to 1.5.  Besides, we have found that the
excited (Efimov) levels for $\lambda<1$ also originate from
virtual levels of the first order that are formed in pairs.
Before a pair of virtual levels appears, there occurs a fusion
of a pair of conjugate resonances of the first order (simple
complex poles of the analytic continuation of the scattering
matrix in the unphysical sheet) resulting in the virtual level
of the second order.

As it will be clear from the further exposition
(see Sect.~\ref{results}), the above-mentioned resonances are not,
generally speaking, genuine resonances of the $^4$He$_3$ trimer
since they are situated outside of the energy domain for which we
can rigorously prove the applicability of the method we are
using for computing the resonances. We will call the resonances
found outside the range of guaranteed applicability of the
method the {\it {\rm(}quasi{\rm)}resonances}.

The paper is organized as follows.

In Sect.~\ref{DiffFS}, we describe the method of search for
resonances in a three--body system on the basis of the Faddeev
differential equations. The idea of the method consists in
calculating the analytic continuation of the component ${\rm
S}_0(z)$ (see formula~(\ref{S0})) of the scattering matrix
corresponding to the ($2+1\to 2+1$) process, in the physical
sheet with the use of these equations.  A particular attention
in this section is paid to the description of the parabolic
domain on the physical sheet where one can analytically continue
the function ${\rm S}_0(z)$ by numerical solving the coordinate
space Faddeev partial equations.  For the potentials we use, the
three--body resonances (including virtual levels) lying in the
unphysical sheet of energy $z$ plane adjoining the physical
sheet along the interval $(\epsilon_d,0)$ are the roots of the
function ${\rm S}_0(z)$ in the physical sheet. We have earlier
employed this method for computing resonances as roots of ${\rm
S}_0(z)$ in the three--nucleon problem~\cite{YaFKM}.

In Sect.~\ref{results}, we first briefly describe the numerical
method we use to solve the $(2+1\to 2+1;\,1+1+1)$ scattering
problem for the $^4$He$_3$ system with going out into the domain
of complex energies. Then we describe the results of our
calculations.

Some notation used throughout the paper is as follows: by ${\Bbb
C}$ we denote the complex plane; $\sqrt{z}$ stands for the main
branch of the function $z^{1/2}$, $\Img\sqrt{z}\geq0$ for any
$z\in{\Bbb C}$; the symbol ${\Bbb R}^2_+$ is used for the
quadrant $x\geq0$, $y\geq0$; by $L_2({\Bbb R}^2_+)$ we
understand the Hilbert space of complex--valued functions which
are integrable on ${\Bbb R}^2_+$ with the absolute value
squared; the symbol $\overline{z}$ stands for the complex number
conjugated to $z$.

\section{Method for search of resonances in a three--body system on
the basis of the Faddeev differential equations}
\label{DiffFS}

\subsection{Faddeev partial differential equations in the case
            of smooth potentials}

In this paper, we will consider the $^4$He$_3$ system
in the state with the total angular momentum $L=0$.

First we consider the case where the interatomic interactions are
described by conventional smooth potentials that include
no hard-core component.
In this case, the angular partial analysis reduces the initial
Faddeev equation for three identical bosons to a system of coupled
two-dimensional integro-differential equations~\cite{MF}
\begin{equation}
\label{FadPart}
   \left[H_{0,l}-z\right]F_l(x,y)=-V(x)\Psi_l(x,y)\,.
\end{equation}
Here, $x,y$ stand for the standard Jacobi variables,
$x\geq0$ and $y\geq0$, and
\begin{equation}
\label{KinPart}
H_{0,l}=-\displaystyle\frac{\partial^2}{\partial x^2}
            -\displaystyle\frac{\partial^2}{\partial y^2}
            +l(l+1)\left(\displaystyle\frac{1}{x^2}
            +\displaystyle\frac{1}{y^2}\right)\,
\end{equation}
for the partial component of the kinetic energy operator.
Functions from the domain of $H_{0,l}$ are assumed to obey the
boundary conditions
\begin{equation}
\label{BCStandard}
      F_{l}(x,y)\left.\right|_{x=0}=0\,, \qquad
      F_{l}(x,y)\left.\right|_{y=0}=0\,.
\end{equation}
which are quite standard when the
expansions over bispherical basis are used. The potential
$V(x)$ is assumed to be central.
In our paper, the energy $z$ can get both real and complex values.
At $L=0$ the partial angular momentum $l$ corresponds both to the dimer
and an additional atom. The momentum $l$ can assume only even values,
$l=0,2,4,\ldots\,.$

The partial wave functions $\Psi_l(x,y)$ are expressed through the Faddeev
partial components $F_l(x,y)$ by the relations
\begin{equation}
\label{FTconn}
         \Psi_l(x,y)=F_l(x,y) + \sum_{l'}\int_{-1}^{+1}
         d\eta\,h_{l l'}(x,y,\eta)\,F_{l'}(x',y')\,
\end{equation}
where
$$
          x'=\sqrt{\displaystyle\frac{1}{4}\,x^2+\displaystyle
    \frac{3}{4}\,y^2-\displaystyle\frac{\sqrt{3}}{2}\,xy\eta}\,,
\qquad
         y'=\sqrt{\displaystyle\frac{3}{4}\,x^2+\displaystyle
   \frac{1}{4}\,y^2+ \displaystyle\frac{\sqrt{3}}{2}\,xy\eta}\,,
$$
and  $-1 \leq{\eta}\leq 1$. The explicit form of the functions
$h_{ll'}$ can be found, e.\,g., in Refs.~\cite{MF,MGL}
(see also~\cite{KMS-JPB}). Here we only deal with a finite number of
equations~(\ref{FadPart}), assuming that
$l\leq l_{\rm max}$ where $l_{\rm max}$ is a certain fixed
even number, $l_{\rm max}\geq 0$. The condition $0\leq l\leq l_{\rm
max}$ is equivalent to the supposition
that the potential $V(x)$ only acts in the two-body
states with $l=0,2,\ldots,l_{\rm max}$. The spectrum of the
Schr\"odinger operator for a system of three identical bosons with such
a potential is denoted by $\sigma_{3B}$.

Its is well known (see, e.\,g., Ref.~\cite{MF}) that if the
potential $V(x)$ is smooth and decreasing as $x\to\infty$
together with its derivatives not slower than
$x^{-3-\varepsilon}$, $\varepsilon>0$, then the asymptotic
conditions as $\rho\rightarrow\infty$ and/or
$y\rightarrow\infty$ for the partial Faddeev components of the
$(2+1\rightarrow 2+1\,;\,1+1+1)$ scattering wave functions%
\footnote{Here we speak about the wave functions usually denoted by
sign ``$(+)$''. The asymptotics of these functions in the total
three-body configuration space ${\Bbb R}^6$ contains, apart from the
incident wave, only the so-called outgoing spherical waves (see, e.g.,
\cite{MF}).}
for  $z=E+{\rm i}0$, $E>0$, read%
\begin{equation}
\label{AsBCPartS}
    \begin{array}{rcl}
      F_l(x,y;z) & = &
      \delta_{l0}\psi_d(x)\left\{\sin(\sqrt{z-\epsilon_d}\,y)
      + \exp({\rm i}\sqrt{z-\epsilon_d}\,y)
      \left[{\rm a}_0(z)+o\left(1\right)\right]\right\} \\
      && +
  \displaystyle\frac{\exp({\rm i}\sqrt{z}\rho)}{\sqrt{\rho}}
                \left[A_l(z,\theta)+o\left(1\right)\right]\,.
    \end{array}
\end{equation}
We assume that the $^4$He$_2$ dimer has an only bound state with
an energy $\epsilon_d$, $\epsilon_d<0$, and wave function
$\psi_d(x)$.  This function is assumed to be
normalized so that for all $x>0$
values of $\psi_d(x)$ are real.  The notations $\rho$,
$\rho=\sqrt{x^2+y^2}$\,, and $\theta$, $\theta=\mathop{\rm
arctg}\displaystyle\frac{y}{x}$\,, are used for the hyperradius
and hyperangle. The coefficient ${\rm a}_0(z)$, $z=E+{\rm i}0$,
for $E>\epsilon_d$ is the elastic scattering amplitude.
The functions $A_l(E+{\rm i}0,\theta)$ provide us, at $E>0$, the
corresponding partial Faddeev breakup amplitudes.  Note that for
$z=E+{\rm i}0$ the correction terms $o(1)$ in coefficients of
outgoing waves $\exp({\rm i}\sqrt{z-\epsilon_d}\,y)$,
$E>\epsilon_d$, and $\exp({\rm i}\sqrt{z}\rho)/\sqrt{\rho}$,
$E>0$, in~(\ref{AsBCPartS}) are of the form, respectively,
$o(y^{-1/2})$ and $o(\rho^{-1/2})$. This property ensures
uniqueness of the solution of the boundary value
problem~(\ref{FadPart}\,--\,\ref{AsBCPartS}) for real scattering
energies $E>\epsilon_d$~\cite{MF}.

The \mbox{$(2+1{\rightarrow}2+1)$} component of
the $s$-wave partial scattering matrix for a system of three
helium atoms is given for real $z=E+{\rm i}0$, $E>\epsilon_d$, by
the expression
\begin{equation}
\label{S0}
{\rm S}_0(z)=1+2{\rm i}{\rm a}_0(z)\,
\end{equation}
while the \mbox{$(2+1\to2+1)$} scattering phases
read
$$
    \delta_0(p)=\frac{1}{2}\,{\rm Im}\,
    \ln{\rm S}_0(\epsilon_d+p^2+{\rm i}0)\,, \quad p>0,
$$
where $p$ stands for the momentum conjugated to the Jacobi
variable~$y$.

\subsection{Holomorphy domains of the Faddeev components
$F_l(z)$ and scattering matrix ${\rm S}_0(z)$}
\label{HolomorphyDomains}

Our goal is to study the analytic continuation of the scattering
matrix ${\rm S}_0(z)$ into the complex plane (the physical
sheet). As it follows from the results of
Refs.~\cite{MotTMFRes,MotMathNachr}, roots of the function ${\rm
S}_0(z)$ in the physical sheet of energy $z$ plane correspond to the
location of the three-body resonances situated in the
unphysical sheet connected with the physical sheet by crossing
the spectral interval $(\epsilon_d,0)$.
This statement is a particular case of more general statements
regarding the three-body resonances
obtained in~\cite{MotTMFRes,MotMathNachr} for the case of two-body
potentials decreasing in the coordinate space not slower than
exponentially. We assume that $V(x)$ is just a
potential which falls off exponentially and, thus,
for all $x\geq0$
\begin{equation}
\label{Vexp}
|V(x)|\leq C\exp(-\mu x)\,,
\end{equation}
with some positive $C$ and $\mu$.  For the sake of simplicity we
even assume sometimes that $V(x)$ is finite, i.\,e.,
\mbox{$V(x)=0$} for $x>r_0$, $r_0>0$.  Looking ahead, we note
that, in fact, in our numerical computations of the $^4$He$_3$ system
at complex energies we make a ``cutoff" of the interatomic
He\,--\,He\,-\,potential at a sufficiently large radius $r_0$.

It is well known that different representations of the same
holomorphic function (for instance, either by a series or by an
integral) allow one to describe this function only in some parts
of its Riemann surface. The
description~\cite{MotTMFRes,MotMathNachr} of the holomorphy
domains for different truncations of the total three-body
scattering matrix in the physical sheet was based on the use of
the Faddeev integral equations in the momentum representation.
In this paper, we make use of the Faddeev equations in the
configuration space. Therefore it is necessary to perform an
investigation, independent of~\cite{MotTMFRes,MotMathNachr}, of
domains in the physical sheet where we can analytically continue
the Faddeev components $F_l(x,y;z)$ and the amplitudes ${\rm
a}_0(z)$ and $A_l(z,\theta)$ just with the use of the
configuration space techniques.

\medskip

Let us list briefly the main results of this investigation
obtained by us for the $^4$He$_3$ system under the
assumption~(\ref{Vexp}). To formulate these results we
distinguish the following three domains in the complex plane
${\Bbb C}$.

\medskip

1$^\circ$. The domain $\Pi^{(\Psi)}$ where the Faddeev components
$F_l(x,y;z)$ (and, hence, the wave functions $\Psi_l(x,y;z)$) can
be analytically continued in $z$ so that the differences
\begin{equation}
\label{FlPhil}
   \Phi_l(x,y;z)=F_l(x,y;z)-
   \delta_{l0}\psi_d(x)\sin(\sqrt{z-\epsilon_d}\,y)
\end{equation}
at $z\in\Pi^{(\Psi)}\setminus\sigma_{3B}$ turn out to be
elements of $L_2({\Bbb R}_+^2)$. The domain $\Pi^{(\Psi)}$ is
described by the inequality
\begin{equation}
\label{PiPsi}
\Img\sqrt{z-\epsilon_d}<{\rm min}\left\{\frac{\sqrt{3}}{2}\mu,\,
\sqrt{3}\sqrt{|\epsilon_d|}\right\}\,.
\end{equation}
For fixed $x,y$ the functions $\Phi_l(x,y;z)$ are continuous in
$z$ up to the rims of the cut along the continuous spectrum
$[\epsilon_d,+\infty)$.
\medskip

2$^\circ$. The domain $\Pi^{(A)}$ where both the elastic scattering
amplitude ${\rm a}_0(z)$ and the Faddeev breakup amplitudes
$A_l(z,\theta)$ can be analytically continued in $z$, $z\not\in\sigma_{3B}$,
provided that the functions $F_l(x,y;z)$
obey the asymptotic formulas~(\ref{AsBCPartS}). This domain
is described by the inequalities
\begin{eqnarray}
\label{PiAmain}
\Img\sqrt{z}+\frac{1}{2}\Img\sqrt{z-\epsilon_d} &<&
\frac{\sqrt{3}}{2}\sqrt{|\epsilon_d|}\,, \\
\label{PiAmu}
\Img\sqrt{z}+\Img\sqrt{z-\epsilon_d} &<&
\frac{\sqrt{3}}{2}\mu\,.
\end{eqnarray}
\medskip

3$^\circ$. And finally, we distinguish the domain $\Pi^{(S)}$,
most interesting for us, where the analytic continuation in $z$,
$z\not\in\sigma_{3B}$, can be only done for the elastic
scattering amplitude ${\rm a}_0(z)$ (and consequently, for the
scattering matrix ${\rm S}_0(z)$); the analytic continuabilty of
the amplitudes $A_l(z,\theta)$ in the whole domain $\Pi^{(S)}$
is not required. The set $\Pi^{(S)}$ is a geometric locus
of points obeying the inequality
\begin{equation}
\label{PiS}
\Img\sqrt{z-\epsilon_d}<
\mathop{\rm min}\left\{\frac{1}{\,\sqrt{3}\,}\,\sqrt{|\epsilon_d|},\,
  \frac{\,\sqrt{3}\,}{2} \mu\right\}\,.
\end{equation}
\medskip

For the domains $\Pi^{(\Psi)}$, $\Pi^{(A)}$, and $\Pi^{(S)}$,
the following chain of inclusions
$$
   \Pi^{(A)}\subset\Pi^{(S)}\subset\Pi^{(\Psi)}.
$$
is valid.

Note that the type~(\ref{PiPsi}) or (\ref{PiS}) condition,
\begin{equation}
\label{Parab}
\Img\sqrt{z-a}<\sqrt{b}, \qquad a\in{\Bbb R},\, b>0,
\end{equation}
is equivalent to the inequality
\begin{equation}
\label{Parabola}
\Real z> a-b+\frac{1}{4b}\,(\Img z)^2\,.
\end{equation}
Therefore, for $\mu\leq2\sqrt{|\epsilon_d|}$ the set $\Pi^{(\Psi)}$
is the domain bounded by the parabola
\begin{equation}
\label{ParabMu}
\Real z>-|\epsilon_d|-\frac{3}{4}\mu^2+\frac{1}{3\mu^2}(\Img z)^2\,.
\end{equation}
For $\mu>2\sqrt{|\epsilon_d|}$ this set coincides with the domain
\begin{equation}
\label{ParabEpsLarge}
   \Real z>-4|\epsilon_d|+\frac{1}{12|\epsilon_d|}(\Img z)^2\,.
\end{equation}
Analogously, if $\mu\leq\frac{2}{3}\sqrt{|\epsilon_d|}$ then
the domain $\Pi^{(S)}$ is described by the inequality~(\ref{ParabMu});
whereas for $\mu>\frac{2}{3}\sqrt{|\epsilon_d|}$ by the inequality
\begin{equation}
\label{ParabEpsSmall}
   \Real z>-\frac{\,4\,}{3}|\epsilon_d|+
\frac{3}{4|\epsilon_d|}(\Img z)^2\,.
\end{equation}
As to the curves bounding the domains~(\ref{PiAmain}) and
(\ref{PiAmu}), we only notice that their order with respect to
the  variables $\Real z$ and $\Img z$ is higher than the second
order. It is easy to check that each of these curves is
connected, symmetric with respect to the $\Real z$ axis and
crosses the latter only once. For the first curve this
intersection occurs at \mbox{$z=\frac{3}{4}\epsilon_d$}, the
slope angle of the tangent at the point of intersection being
independent of $\epsilon_d$, \mbox{$\reduction{\frac{d\Real
z}{d\Img z}}{\Img z=\pm0}=\pm\frac{\sqrt{3}}{2}$}.  As $\Real
z\to+\infty$, the boundaries~(\ref{PiAmain}) and~(\ref{PiAmu})
are asymptotically approximated by the type~(\ref{Parabola})
parabolas with coefficients $a$ and $b$ which can be
computed explicitly.

\medskip

To prove the assertion\,1$^\circ$ concerning the domain $\Pi^{(\Psi)}$
we note that the functions $\Phi_l(x,y;z)$ given by the formulas
(\ref{FlPhil}) satisfy the equations
\begin{equation}
\label{FaddXi}
\left[H_{0,l}+V(x)-z\right]\Phi_l(x,y;z)
 +V(x)\sum_{l'}\int_{-1}^{+1}
 d\eta\,h_{l l'}(x,y,\eta)\,\Phi_{l'}(x',y',z)=
             \chi_l(x,y;z)\,
\end{equation}
where
$$
 \chi_l(x,y;z)=-V(x)\int_{-1}^{+1}
         d\eta\,h_{l0}(x,y,\eta)
      \psi_d(x')\sin(\sqrt{z-\epsilon_d}\,y')\,.
$$
Obviously, for $z\in\Pi^{(\Psi)}$ the functions
$\chi_l(x,y;z)$ fall off exponentially as $\rho\to\infty$. Moreover,
for all the directions $0\leq\theta\leq\pi/2$ the uniform estimate
$$
\chi_l(x,y;z)\mathop{\mbox{\large$=$}}\limits_{\rho\to\infty}
O\biggl(\exp(-\alpha\rho)\biggr)\,
$$
is valid with
$
\alpha={\rm min}\left\{
\frac{\sqrt{3}}{2}\mu-\Img\sqrt{z-\epsilon_d},\,
\frac{\sqrt{3}}{2}\sqrt{|\epsilon_d|}-
\frac{1}{2}\Img\sqrt{z-\epsilon_d}
\right\}\,.
$
Consequently, if the condition~(\ref{PiPsi}) holds then
the inhomogeneous terms $\chi_l(x,y;z)$ considered as functions of
the variables $x$ and $y$ at fixed $z$, are elements of $L_2({\Bbb
R}^2_+)$. At the same time, the vectors $\chi_l(z)$ turn out to be
holomorphic functions of $z\in\Pi^{(\Psi)}$ with respect to
the $L_2({\Bbb R}^2_+)$ norm.

In the problem under consideration, the spectrum of the Faddeev
matrix operator defined by the l.h.s. of Eqs.~(\ref{FaddXi}) and
by the boundary conditions (\ref{BCStandard}) in the Hilbert
space constituted of the vectors
\mbox{$\Phi=(\Phi_0,\Phi_2,\ldots,\Phi_{l_{\rm max}})$},
$\Phi_l\in L_2({\Bbb R}^2_+)$, coincides with the spectrum
$\sigma_{3B}$ of the corresponding three-boson Schr\"odinger
operator with two-body potentials $V(x)$ only acting in the states
with $l=0,2,\ldots,l_{\rm max}$. This means that for any
energy $z\in\Pi^{(\Psi)}$ lying outside of the spectrum
$\sigma_{3B}$, the inhomogeneous system~(\ref{FaddXi}) is
uniquely solvable in the class of the functions $\Phi_l(z)\in
L_2({\Bbb R}^2_+)$, $l=0,2,\ldots,l_{\rm max}$.  Since outside
of the set $\sigma_{3B}$ the resolvent of the Faddeev operator
is a holomorphic operator-valued function of the variable $z$,
each of the components $\Phi_l(z)$ of the solution of
Eqs.~(\ref{FaddXi}) also is a holomorphic function of
$z\in\Pi^{(\Psi)}\setminus\sigma_{3B}$. The bound-state energies
of the three-boson system under consideration turn
out to be poles of the first order for $\Phi_l(z)$.  Thus, the
Faddeev partial components $F_l(x,y;z)$ admit the analytic
continuation in $z$ in the form~(\ref{FlPhil}) into the domain
$\Pi^{(\Psi)}\setminus\sigma_{3B}$.

\medskip

The proof of the assertions of \,2$^\circ$ and 3$^\circ$ regarding
the domains $\Pi^{(S)}$ and $\Pi^{(A)}$ is rather cumbersome.
This is why we here only outline its main steps. Note that the
proof is based on the integral equations method and it
is quite standard (see, for instance, Ref.~\cite{MF}, Chapter~V).
First, the equations~(\ref{FaddXi}) are rewritten in the form of
the Faddeev partial integral equations. To do this, it suffices to
reverse the operators  $H_{0,l}+V-z$ in (\ref{FaddXi}). Since
the variables $x$ and $y$ in $H_{0,l}$ are separated,
the kernels $R^{(l)}_v(X,X';z)$, $X=\{x,y\}$, $X'=\{x',y'\}$, of
the respective resolvents $R^{(l)}_v(z)=(H_{0,l}+V-z)^{-1}$
are explicitly expressed in terms of the two-body problem.
Analytic properties in the variable $z$ and coordinate
asymptotics of the kernels $R^{(l)}_v(X,X';z)$ are well known
(see Ref.~\cite{MF}, Chapters~IV and~V).  Iterations first ``improve"
and then stabilize the asymptotic properties of the iterated kernels
and inhomogeneous terms of the Faddeev equations. (In the case under
consideration, this stabilization requires only three
iterations.) Further, it turns out that, for
$z\not\in[\epsilon_d,+\infty)$, the iterated kernels are
represented by sums of exponentially decreasing terms admitting,
in certain domains of the configuration space, an explicit
asymptotic factorization with respect to
$X$ and $X'$. Since we are working in the domain where
$\chi_l(z)\in L_2({\Bbb R}^2_+)$, the corresponding asymptotic
factors of these terms, along with the asymptotics of the iterations
of the inhomogeneous term, determine the coordinate asymptotics
of the functions $\Phi_l(x,y;z)$. Therefore, finally we are able
to determine the geometric locus of the points $z$ in the
complex plane for which there exists a (non-empty) set in the
configuration space such that the leading term of the coordinate
asymptotics of the function $\Phi_l(x,y;z)$ in this set
represents a term of the form ${\rm a}_0(z)\psi_d(x)\exp({\rm
i}\sqrt{z-\epsilon_d}\,y)$, and thus, for these $z$ the
scattering matrix ${\rm S}_0(z)$ is well defined. This geometrical
locus is just the domain $\Pi^{(S)}$.  In this domain, as
$y\to\infty$ and/or $\rho\to\infty$, the functions
$\Phi_l(x,y;z)$ admit the asymptotic representation
\begin{eqnarray}
\label{PhiPiS}
\Phi_l(x,y;z) &=&
   \delta_{l0}\psi_d(x)\left\{\exp({\rm i}\sqrt{z-\epsilon_d}\,y)
\left[{\rm a}_0(z)+o(1)\right]+{\rm     f}_0(y;z)\right\} \\
\nonumber
&+&  \displaystyle\frac{\exp({\rm i}\sqrt{z}\rho)}{\sqrt{\rho}}
     \left[A_l(z,\theta)+o(1)\right] + {\rm F}_{1,l}(x,y;z)
\end{eqnarray}
with
\begin{equation}
\label{f01l}
{\rm  f}_0(y;z)\mathop{\mbox{\large$=$}}\limits_{y\to\infty}
O\left({\rm e}^{-\alpha_0(z)y}\right)\qquad\mbox{and}\qquad
{\rm  F}_{1,l}(x,y;z)\mathop{\mbox{\large$=$}}\limits_{\rho\to\infty}
O\left({\rm e}^{-\alpha_1(z)\rho}\right)\,
\end{equation}
where
\begin{eqnarray}
\label{alfa0}
\alpha_0(z)&=&\frac{\sqrt{3}}{2}\,\sqrt{|\epsilon_d|}-\frac{\,1\,}{2}\,
\Img\sqrt{z-\epsilon_d}\,, \\
\label{alfa1}
\alpha_1(z)&=& \mathop{\rm min}\left\{\alpha_0(z)\,,\,\,
\frac{\sqrt{3}}{2}\,\mu-\Img\sqrt{z-\epsilon_d}\,,\,
\Img\sqrt{z}\right\}\,.
\end{eqnarray}
In a parabolic neighborhood of the  $y$-axis, the functions
${\rm F}_{1,l}(x,y;z)$ are also subjected to the asymptotic estimates
\begin{equation}
\label{F1nu}
{\rm F}_{1,l}(x,y;z)
\mathop{\mbox{\large$=$}}\limits_{
\begin{array}{c} y\to\infty \\ x<y^\nu \end{array}
}
O\biggl(\exp(-\alpha_0(z)y)\biggr)\,
\end{equation}
where $\nu$ is an arbitrary fixed number smaller than
unity, $\nu<1$.

As to the domain $\Pi^{(A)}$, the leading asymptotic term of
each of the functions ${\rm F}_{1,l}(x,y;z)$ for $z\in\Pi^{(A)}$
is a spherical wave $\exp({\rm i}\sqrt{z}\rho)/\sqrt{\rho}$ with
the amplitude $A_{1,l}(\theta)$ being a differentiable function
of the angle $\theta$.  Therefore, for $z\in\Pi^{(S)}$ the term
${\rm F}_{1,l}(x,y;z)$ in the r.h.s. of the formula~(\ref{PhiPiS})
can be added to the asymptotic term with a spherical wave
preceding ${\rm F}_{1,l}$. In the domain $\Pi^{(S)}$, and hence,
in a narrower domain $\Pi^{(A)}$ the condition
$
{\rm f}_0(y;z)\mathop{\mbox{\large$=$}}\limits_{y\to\infty}
o\biggl(\exp({\rm i}\sqrt{z-\epsilon_d}\,y)\biggr)\,
$
holds. Consequently, for $z\in\Pi^{(A)}$, the Faddeev components
$F_l(x,y;z)$ do obey the standard asymptotic conditions
like (\ref{AsBCPartS}).

Therefore, for any $\nu<1$ the dominant term of the asymptotics
of the function $\Phi_0(x,y;z)$,
$z\in\Pi^{(S)}\setminus\sigma_{3B}$, in the domain $x<y^\nu$
reads as ${\rm a}_0(z)\psi_d(x)\exp({\rm
i}\sqrt{z-\epsilon_d}\,y)$ as $y\to\infty$.  This means that,
for $z\in\Pi^{(S)}\setminus\sigma_{3B}$, it is always possible
by solving the equations~(\ref{FadPart}) to separate
explicitly the elastic scattering amplitude ${\rm a}_0(z)$ and,
thus, to construct the analytic continuation of the scattering
matrix ${\rm S}_0(z)$.

Outside of the domain $\Pi^{(S)}$ the numerical construction of
${\rm S}_0(z)$ by solving the Faddeev differential equations is,
in general, impossible since for $x<y^\nu$ and $\nu<1$ both
functions ${\rm f}_0(y;z)$ and ${\rm F}_{1,0}(x,y;z)$,
$z\not\in\Pi^{(S)}$, include terms decreasing
slower than $\exp({\rm i}\sqrt{z-\epsilon_d}\,y)$
as $y\to\infty$.

\subsection{The partial Faddeev differential equations in the
case of potentials with hard core}

In the case of potentials with hard core, the partial Faddeev
differential equations for a system of three identical bosons at
$L=0$ acquire the form
\begin{equation}
\label{FadPartCor}
   \left[H_{0,l} -z\right]F_l(x,y)=\left\{
            \begin{array}{cl} -V(x)\Psi_l(x,y), & x>c \\
                    0,                  & x<c\,,
\end{array}\right.
\end{equation}
where $c$, $c>0$, is the core size. The partial wave functions
$\Psi_l(x,y)$ are expressed via Faddeev partial
components $F_l(x,y)$ by the formulas~(\ref{FTconn}).
The components $F_l(x,y)$ satisfy the standard boundary conditions
(\ref{BCStandard}). The two-body central potential $V(x)$
acts only beyond the core domain, i.\,e. only where $x>c$.
We assume as before that $V(x)$ falls off not
slower than exponentially  as \mbox{$x\to\infty$} and, hence, it
satisfies the condition (\ref{Vexp}) for some $C>0$ and
$\mu>0$.

A main difference between the model with hard core and
those with smooth potentials is that the functions $F_l(x,y)$ in
this model satisfy the auxiliary boundary conditions
\begin{equation}
\label{BCCorePart}
       F_l(c,y) + \sum_{l'}\int_{-1}^{+1}
       d\eta\,h_{l l'}(c,y,\eta)\,F_{l'}(x',y')=0\,
\end{equation}
requiring that the wave functions $\Psi_{l}(x,y)$ vanish on the
boundary  $x=c$ of the core domain. It can be shown that in fact
the conditions (\ref{BCCorePart}) force the wave functions
(\ref{FTconn}) to vanish also inside the core domain at all energies
except for a certain countable set of real values of
$z$ (see Ref.~\cite{KMS-JPB} and references therein).

Asymptotic conditions for the partial Faddeev components
$F_l(x,y;z)$ of the $(2+1\rightarrow 2+1\,;\,1+1+1)$ scattering
wave functions as $\rho\rightarrow\infty$ and/or
$y\rightarrow\infty$ are again of the form (\ref{AsBCPartS}).
The only difference is that the dimer wave function $\psi_d(x)$
is considered as zero in the core domain, i.\,e. for $0\leq
x\leq c$.

In the hard-core model, all the assertions of Sect.
\ref{HolomorphyDomains} regarding the holomorphy domains of
the functions $\Phi_l(x,y;z)$ and the scattering
matrix ${\rm S}_0(z)$ in $z$ still hold true.

\subsection{Resonances and virtual levels as roots of the
scattering matrix ${\rm S}_0(z)$ in the physical sheet}

We have already noticed that the roots of ${\rm S}_0(z)$ in the
physical sheet of energy $z$ plane correspond to the location of
the three-body resonances in the unphysical sheet adjoining the
physical sheet along the spectral interval $(\epsilon_d,0)$. In the
case under consideration, this statement is an immediate
consequence of the unitarity of the scattering matrix ${\rm
S}_0(z)$ for $z=E+{\rm i}0$, $\epsilon_d\leq E\leq0$,
\begin{equation}
\label{S0unitar}
      {\rm S}_0(E+{\rm i}0)\,\overline{{\rm S}_0(E+{\rm i}0)}=1\,.
\end{equation}
Indeed, as we have established, the functions $\Phi_l(x,y;z)$
are holomorphic functions of
$z\in\Pi^{(\Psi)}\setminus\sigma_{3B}$. Since the boundary value
problem~(\ref{FadPart}\,--\,\ref{AsBCPartS}) is uniquely
solvable, one easily verifies that the boundary values
\mbox{$\Phi_l(x,y;E+{\rm i}0)$} and \mbox{$\Phi_l(x,y;E-{\rm
i}0)$} for these functions on the rims of the cut along
\mbox{$[\epsilon_d,+\infty)$} are related to each other as
\begin{equation}
\label{Phi-pm}
\Phi_l(x,y;E+{\rm i}0)=-\overline{\Phi_l(x,y;E-{\rm i}0)}\,
\end{equation}
since, on the one hand, their asymptotics~(\ref{PhiPiS})
as $y\to\infty$ and/or $\rho\to\infty$ has the same structure and,
on the other hand,
$$
\chi_l(x,y;E+{\rm i}0)=-\chi_l(x,y;E-{\rm i}0)=
-\overline{\chi_l(x,y;E-{\rm i}0)}\,,
$$
since
$$
\sin(\sqrt{E-\epsilon_d+{\rm i}0}\,y)=
-\sin(\sqrt{E-\epsilon_d-{\rm i}0}\,y)=
-\overline{\sin(\sqrt{E-\epsilon_d-{\rm i}0}\,y)}\,.
$$
Consequently,
\begin{equation}
\label{aadj}
{\rm a}_0(E+{\rm i}0)=-\overline{{\rm a}_0(E-{\rm i}0)}\,
\end{equation}
and
\begin{equation}
\label{Spm}
     {\rm S}_0(E+{\rm i}0)=\overline{{\rm S}_0(E-{\rm i}0)}\,,
\qquad E>\epsilon_d\,.
\end{equation}
Therefore, it follows from Eq.~(\ref{S0unitar})
that for $\epsilon_d\leq E\leq0$
$$
   {\rm S}_0(E+{\rm i}0)=[{\rm S}_0(E-{\rm i}0)]^{-1}
\qquad\mbox{and}\qquad
   {\rm S}_0(E-{\rm i}0)=[{\rm S}_0(E+{\rm i}0)]^{-1}\,.
$$
This means that the function ${\rm S}_0(E+{\rm i}0)$ is
continued through the cut $[\epsilon_d,0]$ into the domain $\Img
z<0$ as ${\rm S}_0^{-1}(z)$. In a similar manner, ${\rm
S}_0(E-{\rm i}0)$ is continued into the domain $\Img z>0$, again
as ${\rm S}_0^{-1}(z)$. All this signifies that the scattering
matrix ${\rm S}_0(z)$ admits analytic continuation at least into
the domain $\Pi^{(S)}$ of the unphysical energy sheet connected
with the physical sheet by crossing the interval
$[\epsilon_d,0]$, the value of the continued function ${\rm
S}_0(z)$ at $z\in\Pi^{(S)}$ in the unphysical sheet coinciding
with the value of ${\rm S}_0^{-1}(z)$ at the same $z$ but in the
physical sheet.

Recall that those points $z$ on unphysical sheets are called resonances
where the analytically continued scattering matrix possesses poles.
The resonances $z$ with zero imaginary part $\Img z=0$ and
$\Real z<\epsilon_d$ are called the virtual levels.

Thus, we have here presented a simple proof of the fact that
the resonances including the virtual levels corresponding to poles of
the analytic continuation of the scattering matrix ${\rm
S}_0(z)$ in the unphysical sheet connected with the physical one
by crossing the spectral interval $[\epsilon_d,0]$ are the roots
of this matrix in the physical sheet. At the same time, the
poles of the function ${\rm S}_0(z)$ in the physical sheet
correspond to bound states of the three-boson system under
consideration.

Concluding the subsection, we note that it follows from
Eq.~(\ref{aadj}) that \mbox{${\rm a}_0(z)=-\overline{{\rm
a}_0(\overline{z})}$} and, hence,
\begin{equation}
\label{S0Sym}
\overline{{\rm
S}_0(z)}={{\rm S}_0(\overline{z})}
\end{equation}
for any $z\in\Pi^{(S)}$. This means that the roots of the function
${\rm S}_0(z)$ are situated symmetrically with respect to
the real axis.

\section{Numerical method and results of computations}
\label{results}

In the present work we make use of the Faddeev equations
(\ref{FadPartCor}) considered together with the boundary
conditions (\ref{BCStandard}), (\ref{AsBCPartS}) and
(\ref{BCCorePart}) to calculate the values of the $^4$He$_3$
scattering matrix ${\rm S}_0(z)$ in the physical sheet. We
search for the resonances including the virtual levels as roots
of ${\rm S}_0(z)$ and for the bound-state energies as positions
of poles of ${\rm S}_0(z)$.  All the results
presented below are obtained for the case $l_{\rm max}=0$.

In all our calculations, \mbox{$\hbar^2/m=12.12$~K\,\AA$^2$.}
As the interatomic He\,--\,He\,-\,interaction we employed
the widely used semiempirical potential HFD-B constructed
by R.\,A.\,Aziz and co-workers~\cite{Aziz87}. This potential
is of the form
\begin{equation}
\label{Aziz1-2}
 V_{\rm HFD-B}(x)=\varepsilon \left \{ A\exp(-\alpha\zeta
	+\beta\zeta^2 ) -\left [ \frac{C_6}{\zeta^6}
	+ \frac{C_8}{\zeta^8} +
 \frac{C_{10}}{\zeta^{10} } \right ]F(\zeta)  \right \}\,
\end{equation}
where
$\zeta=x/r_m$. The function
$F(\zeta)$ reads
$$
 F(\zeta)=\cases { \exp{ \left [ -\left( D/\zeta
 -1 \right) \right]^2},
    & \mbox{if  $\zeta\leq D$}   \cr
1,  & \mbox{if  $\zeta >  D $}\,. }
$$
For completeness the parameters of the potential HFD-B
are given in Table~I.

The value of the parameter $c$ (the core ``diameter" of
particles) is chosen to be so small that its further decrease
does not appreciably influence the dimer binding energy
$\epsilon_d$ and the energy of the trimer ground state
$E_t^{(0)}$. Unlike papers~\cite{KMS-PRA,MSK-CPL,KMS-JPB}, where
$c$ was taken to be equal 0.7\,{\AA}, now we take
$c=1.3$\,{\AA}.  We have found that such a value of
$c$ provides at least six reliable figures of
$\epsilon_d$ and three figures of $E_t^{(0)}$.

Since the statements of Sect.~\ref{DiffFS} are valid, generally
speaking, only for the potentials decreasing not slower than
exponentially, we cut off the potential HFD-B setting $V(x)=0$
for $x>r_0$. We have established that this cutoff for
\mbox{$r_0\gtrsim 95$\,{\AA}} provides the same values of
$\epsilon_d$ ($\epsilon_d=-1.68541$\,mK), $E_t^{(0)}$
($E_t^{(0)}=-0.096$\,K) and phases $\delta_0(p)$
which were obtained in our earlier calculations
\cite{KMS-PRA,MSK-CPL,KMS-JPB} performed with the potential
HFD-B. Comparison of these results with results of other
researchers can be found in
Refs.~\cite{KMS-PRA,MSK-CPL,KMS-JPB}. In all the calculations of
the present work we take \mbox{$r_0=100$\,{\AA}}.  Note that if
the formulas from Sect. \ref{DiffFS} including the parameter
$\mu$ are used for finite potentials, one should set
$\mu=+\infty$.

Before making numerical approximation of the system of equations
(\ref{BCStandard}), (\ref{FadPartCor}), (\ref{BCCorePart}) at
$l_{\rm max}=0$ we rewrite it in terms of a new unknown function
$\Phi_0(x,y;z)$ that is expressed via the Faddeev component
$F_0(x,y;z)$ by the relation~(\ref{FlPhil}).
Note that for $z\in\Pi^{(\Psi)}\setminus\sigma_{3B}$ the function
$\Phi_0(x,y;z)$ is square integrable in
$x,y$ (see Sect.~\ref{HolomorphyDomains}). Therefore,
this function is uniquely determined by the asymptotic condition
\begin{equation}
\label{PhiAs0}
\Phi_0(x,y;z)\mathop{\longrightarrow}
\limits_{\rho\to\infty}0\,
\end{equation}
that can be easily approximated and programmed.  One could, for
instance, require
$\reduction{\Phi_0(x,y;z)}{\sqrt{x^2+y^2}=\rho_{\rm max}}=0$ at
a sufficiently large $\rho_{\rm max}$ and look for a numerical
solution of the system (\ref{BCStandard}), (\ref{FadPartCor}),
(\ref{BCCorePart}) satisfying this condition. Further, for
$z\in\Pi^{(S)}$, one could, going sufficiently far from
$\rho_{\rm max}$ into the domain of smaller (but nevertheless,
providing the asymptotics~(\ref{PhiPiS})) values of $\rho$,
separate the elastic scattering amplitude ${\rm a}_0(z)$,
putting, e.\,g., ${\rm a}_0(z)\approx\Phi_0(x,y;z)\exp(-{\rm
i}\sqrt{z-\epsilon_d}\,y)$, where the value of $x$ corresponds
to the maximum of the function $\psi_d(x)$.  Such an approach
is, however, not effective in view of a relatively slow decrease
of the exponentials $\exp(-\sqrt{|\epsilon_d|}\,x)$ and
$\exp(-\Img\sqrt{z-\epsilon_d}\,y)$ as well as of the function
$\exp(-\Img\sqrt{z}\,\rho)$ in the energy domain of interest for
us in $\Pi^{(S)}$. For a proper approximation of the
condition~(\ref{PhiAs0}), very large values of $\rho_{\rm max}$
are to be taken.  This is just a reason why one should take into
account the asymptotics of the function $\Phi_0(x,y;z)$ as
$x\to\infty$ and/or $y\to\infty$. Though the asymptotic
formula~(\ref{AsBCPartS}) only holds for $z\in\Pi^{(A)}$, we
employ it also for $z\in\Pi^{(S)}\setminus\Pi^{(A)}$. Indeed,
when $z\in\Pi^{(S)}\setminus\Pi^{(A)}$, the leading term of the
asymptotics of $\Phi_0(x,y;z)$ as $y\to\infty$ and $x<y^\nu$,
$\nu<1$, is given by the same expression \mbox{${\rm
a}_0(z)\exp({\rm i}\sqrt{z-\epsilon_d}\,y)$} (see Sect.
\ref{HolomorphyDomains}) as in Eq.~(\ref{AsBCPartS}). Outside of
the parabola $x<y^\nu$, it suffices to require the condition
(\ref{PhiAs0}) to be satisfied. The presence, in
Eq.~(\ref{AsBCPartS}), of the spherical wave \mbox{$\exp({\rm
i}\sqrt{z}\,\rho)/\sqrt{\rho}$} does not contradict this
requirement. Therefore, the use of asymptotic condition
(\ref{AsBCPartS}) is justified even if
$z\in\Pi^{(S)}\setminus\Pi^{(A)}$.

A detailed description of the numerical method we use is
presented in Ref.~\cite{KMS-JPB}. Here we only mention main
steps of the computational scheme~\cite{KMS-JPB} helpful for
understanding our results.

When solving the boundary-value problem (\ref{BCStandard}),
(\ref{AsBCPartS}), (\ref{FadPartCor}), (\ref{BCCorePart})
written in terms of the function $\Phi_0$, we carry out its
finite-difference approximation in polar coordinates $\rho$ and
$\theta$.  The grid is chosen in such a way that
the points of intersection of arcs
$\rho=\rho_i$, $i=1,2,\ldots, N_\rho$, and rays
$\theta=\theta_j$, $j=1,2,\ldots, N_\theta$, with the line $x=c$
turn out automatically to be its knots. The $\rho_i$ points are chosen
according to the formulas
\begin{eqnarray}
\nonumber
\rho_i &=&\frac{i}{N_c^{(\rho)}+1}\, c,
\quad i=1,2,\ldots,N_c^{(\rho)},\\
\nonumber
		\rho_{i+N_c^{(\rho)}}& = & \sqrt{c^2 + y_i^2}, \quad
	     	i=1,2,\ldots,N_\rho-N_c^{(\rho)},
\end{eqnarray}
where $N_c^{(\rho)}$ stands for the number
of arcs inside the core domain and
$$
		y_i = f(\tau_i)\sqrt{\rho^2_{N_\rho}-c^2}, \quad
		\tau_i = \frac{i}{N_\rho-N_c^{(\rho)}}.
$$
The nonlinear monotonously increasing function
$f(\tau)$,
$0\leq\tau\leq 1$, satisfying the conditions $f(0)=0$ and
$f(1)=1$ is chosen in the form
$$
		f(\tau)=\left\{
		\begin{array}{lcl} \alpha_0\tau & , & \tau\in[0,\tau_0]\\
		\alpha_1\tau+\tau^\nu &,& \tau\in(\tau_0,1]
		\end{array}
		\right..
$$
The values of $\alpha_0$, $\alpha_0\geq 0,$ and $\alpha_1$,
$\alpha_1\geq 0,$ are determined via $\tau_0$ and $\nu$ from the
continuity condition for $f(\tau)$ and its derivative at the
point $\tau_0$. As a rule, we took values of $\tau_0$ within
0.1 and 0.2.  The value of the power $\nu$ depends on the cutoff
radius $\rho_{\rm max}=\rho_{N_\rho}=50\,$---$\,4100\,{\AA}$ its
range being within 2 and 4 in our calculations.

The knots $\theta_j$ at $j=1,2,\ldots,N_\rho-N_c^{(\rho)}$
are taken according to $\theta_j=\mathop{\rm arctg}(y_j/c)$.
The rest knots $\theta_j$, $j=N_\rho-N_c^{(\rho)}+1,
\ldots,N_\theta,$ are chosen equidistantly.
Such a choice of the grid is prescribed by the need
to have a higher density of points
in the domain where the functions $\Phi_l(x,y;z)$ are most
rapidly changing, i.\,e. for small values of $\rho$ and/or $x$
and lower in the asymptotic domain.
In this work, we used the grids of dimension
\mbox{$N_\theta=N_\rho=$\,600\,---\,1000}.  The number of the
last arc knots in $\theta$ lying in the core domain was usually
equal to $N_c^{(\rho)}=5$.

The finite-difference approximation of the integro-differential
equations (\ref{FadPartCor}) and boundary conditions
(\ref{BCStandard}), (\ref{BCCorePart}) for $l_{\rm max}=0$
reduces the problem to a system of $N_\rho N_\theta$ linear
algebraic equations.  The finite-difference equations
corresponding to the arc $i=N_\rho$ include initially the values
of the unknown function $\Phi(x,y;z)$ from the arc $i=N_\rho+1$.
To eliminate them, we express these values through the values
of $\Phi(x,y;z)$ on the arcs $i=N_\rho$ and $i=N_\rho-1$ by
using the asymptotic formula (\ref{AsBCPartS}), just in the
manner described in the concluding part of Appendix A of
Ref.~\cite{KMS-JPB}. In \cite{KMS-JPB}, this approach was only used
for computing the energies of bound states. Now we extend
it also on the scattering problem. (Note that the
formulas (A10) and (A11) in \cite{KMS-JPB} related to the
described approach contain misprints.
The values $C^-_{N_\rho}$ in these
formulas should be replaced with inverse values $1/C^-_{N_\rho}$.) The
matrix of the resultant system of equations has a
block-three-diagonal form (see Ref.~\cite{KMS-JPB}, Appendix A). Every
block has the dimension $N_\theta\times N_\theta$ and consists of the
coefficients standing at unknown values of
the function $\Phi(x,y;z)$ in the grid knots belonging to a certain
arc $\rho=\rho_i$.  The main diagonal of the matrix
consists of $N_\rho$ such blocks.

In contrast to \cite{KMS-PRA,MSK-CPL,KMS-JPB}, in the present
paper we solve the block-three-diagonal algebraic system on the
basis of the matrix sweep method. This allows us to dispense
with writing the system matrix on the hard drive and to carry
out all the operations related to its inversion immediately in
RAM. Besides, the matrix sweep method reduces almost by one
order the computer time required for computations on the grids of
the same dimensions as in~\cite{KMS-PRA,MSK-CPL,KMS-JPB}.

We searched for the resonances (roots of the function ${\rm
S}_0(z)$ on the physical sheet) and bound-state energies
(roots of the function ${\rm S}_0^{-1}(z)$ for real
$z<\epsilon_d$) of the helium trimer by using the complex
version of the secant method. Within this method, the
approximation $z_n$ to a root of a holomorphic function
$f(z)$ is constructed from the two previous approximations $z_{n-1}$
and $z_{n-2}$ according to the formula
$$
     z_n=z_{n-1}-\frac{f(z_{n-1})(z_{n-1}-z_{n-2})}
     {f(z_{n-1})-f(z_{n-2}))}\,.
$$

As the relationship (\ref{S0Sym}) implies the symmetry of properties of
the scattering matrix ${\rm S}_0(z)$ with respect to the real
axis, we performed all the calculations for ${\rm S}_0(z)$ only
for $\Img z\geq 0$ (except the tests of the code). We start with
a study of graph surfaces of the real and imaginary
parts of the scattering matrix
${\rm S}_0(z)$ in the domain of its holomorphy
$\Pi^{(S)}\setminus\sigma_{3B}$. The root lines of the functions
$\Real{\rm S}_0(z)$ and $\Img{\rm S}_0(z)$ obtained in the case
of the grid parameters \mbox{$N_\theta=N_\rho=600$} and
\mbox{$\rho_{\rm max}=600$}\,{\AA} are depicted in Fig.~1. Both
resonances (roots of ${\rm S}_0(z)$) and bound-state energies (poles of
${\rm S}_0(z)$) of the $^4$He trimer are associated with the
intersection points of the curves $\Real{\rm S}_0(z)=0$ and
$\Img{\rm S}_0(z)=0$. When the roots or poles are simple, these
curves intersect each other at the right angle. Note that for
real $z\leq\epsilon_d$ the function ${\rm S}_0(z)$ is real and,
thus, $\Img{\rm S}_0(z)=0$.  In Fig.~1, along with the root
lines we also plot the boundaries of the domains $\Pi^{(S)}$,
$\Pi^{(A)}$ and $\Pi^{(\Psi)}$.  One can observe that a ``good"
domain $\Pi^{(S)}$ includes none of the points of intersection
of the root lines $\Real{\rm S}_0(z)=0$ and $\Img{\rm
S}_0(z)=0$. Nevertheless, as we will see
below, the going beyond this domain is of an interest,
even though the asymptotic formula (\ref{AsBCPartS}) is not
valid for $z\in{\Bbb C}\setminus\Pi^{(S)}$ and the function
${\rm S}_0(z)$ calculated there cannot be interpreted as the
scattering matrix.  The caption for Fig.~1 points out positions
of the four ``resonances", the roots of ${\rm S}_0(z)$, found
immediately beyond the boundary of the domain  $\Pi^{(S)}$.  As
one could expect, the values of the function ${\rm S}_0(z)$ at
$z\in{\Bbb C}\setminus\Pi^{(S)}$ and positions of its roots in
${\Bbb C}\setminus\Pi^{(S)}$ turn out to be unstable and
strongly depend on the value of the cutoff radius $\rho_{\rm
max}$, whereas the dependence on the number of knots is weak. In
particular, for $\rho_{\rm max} =400$\,{\AA}, a
(quasi)resonance, closest to the real axis, is situated at the
point \mbox{$(-1.95+{\rm i}\,1.81)$\,mK,} if
\mbox{$N_\theta=N_\rho=300$,} at the point \mbox{$(-1.90+{\rm
i}\,1.85)$\,mK,} if \mbox{$N_\theta=N_\rho=520$,} and at the
point \mbox{$(-1.89+{\rm i}\,1.86)$\,mK} if
\mbox{$N_\theta=N_\rho=800$.} The same (quasi)resonance in
Fig.~1 (calculated for \mbox{$\rho_{\rm max}=600$\,\AA}) is
situated at the point $(-2.34+{\rm i}\,0.97)$~mK. If
$N_\theta=N_\rho=600$ is fixed, the increase of $\rho_{\rm
max}$ up to 800\,{\AA} shifts this point to the point
$(-2.44+{\rm i}\,0.65)$\,mK.

All the aforesaid regarding the instability of the function
${\rm S}_0(z)$ values and positions of its roots beyond the domain
$\Pi^{(S)}$ bears no relation to its pole at the point
\mbox{$z=E_t^{(1)}=-2.46$\,mK}, corresponding to a trimer
excited-state energy, even though this energy
does not belong to $\Pi^{(S)}$.
The point is that the position of the pole of ${\rm
S}_0(z)$ is only determined by the position of the root of
the determinant of the linear algebraic system we solve, whereas the
inhomogeneous term of the system plays no role. Therefore, the
search for the poles of the grid function ${\rm S}_0(z)$ is
equivalent to the search for the binding energies of the trimer.
The grids we have used turn out to be quite sufficient for this
purpose. The convergence of our results for $E_t^{(1)}$ with
respect to the parameters $N_\theta, N_\rho, \rho_{\rm max}$ and
their accuracy can be judged from the values of the difference
$\epsilon_d-E_t^{(1)}$ obtained with different grids and shown in
Table~II.

We would like to stress that we do not consider the roots of
function ${\rm S}_0(z)$ drawn in Fig.~1 as genuine resonances
for the $^4$He$_3$ system since they are situated beyond the domain
$\Pi^{(S)}$ where the applicability of our method is proved.  We
should rather consider them as artifacts of the method. However it is
remarkable that the ``true" (i.\,e., getting inside
$\Pi^{(S)}$) virtual levels and then the energies of the excited
(Efimov) states appear just due to these (quasi)resonances when
the potential $V(x)$ is weakened. This is the object of our
further consideration.

Following~\cite{EsryLinGreene,Gloeckle,KMS-JPB},
instead of the initial potential $V(x)=V_{\rm HFD-B}(x)$,
we will consider the potentials
$$
          V(x)=\lambda\cdot V_{\rm HFD-B}(x).
$$
To establish the mechanism of formation of new excited
states in the $^4$He trimer, we have first calculated the scattering
matrix ${\rm S}_0(z)$ for $\lambda<1$. In Table~III
for some values of $\lambda$ from the interval between 0.995
and 0.975, we present the positions of roots and poles of ${\rm S}_0(z)$,
we have obtained at real $z<\epsilon_d(\lambda)$. We have found
that for a value of $\lambda$ slightly smaller than $0.9885$, the
(quasi)resonance closest to the real axis (see
Fig.~1) gets on it and transforms into a virtual level
(the root of ${\rm S}_0(z)$) of the second order corresponding to
the energy value where the graph of ${\rm S}_0(z)$,
$z\in{\Bbb R}$, $z<\epsilon_d$, is tangent to the axis $z$. This virtual
level is preceded by the (quasi)resonances
\mbox{$z=(-1.04+{\rm i}\,0.11)$}\,mK
\mbox{$(z/|\epsilon_d|=-1.58+{\rm i}\,0.168)$} for $\lambda=0.989$
and \mbox{$z=(-0.99+{\rm i}\,0.04)$}\,mK
\mbox{$(z/|\epsilon_d|=-1.59+{\rm i}\,0.064)$} for
$\lambda=0.9885$.  The originating virtual level is of the second order
since simultaneously with the root of the function ${\rm
S}_0(z)$, also the conjugate root of this function gets on the
real axis. With a subsequent decrease of $\lambda$ the virtual
level of the second order splits into a pair of the virtual
levels $E_t^{(2)*}$ and $E_t^{(2)**}$, $E_t^{(2)*}<E_t^{(2)**}$
of the first order which move in opposite directions. A
characteristic behavior of the scattering matrix ${\rm S}_0(z)$
when resonances transform into virtual levels is shown in
Fig.~2. The virtual level $E_t^{(2)**}$ moves towards the
threshold $\epsilon_d$ and ``collides'' with it at
$\lambda<0.98$. For $\lambda=0.975$ the function ${\rm S}_0(z)$
has no longer the root corresponding to $E_t^{(2)**}$. Instead
of the root, it acquires a new pole corresponding to the second
excited state of the trimer with the energy $E_t^{(2)}$. Note
that though the virtual levels $E_t^{(2)*}$ and $E_t^{(2)**}$
appear beyond the domain $\Pi^{(S)}$, already at $\lambda=0.985$
the point $E_t^{(2)**}$ turns out to be inside this domain.
Therefore, it should be considered as a ``true'' virtual level
of the trimer.  We expect that the subsequent Efimov levels
originate from the virtual levels just according to the same
scheme as the level $E_t^{(2)}$ does.

The other purpose of the present investigation is to determine
the mechanism of disappearance of the excited state of the
helium trimer when the two-body interactions become stronger
owing to the increasing coupling constant $\lambda>1$.  It
turned out that this disappearance proceeds just according to
the scheme of the formation of new excited states; only the
order of occurring events is inverse.

The results of our computations of the energy $E_t^{(1)}$ when
$\lambda$ changes from 1.05 to 1.17 are given in Table~IV.  In
the interval between $\lambda=1.17$ and $\lambda=1.18$ there
occurs a ``jump" of the level $E_t^{(1)}$ on the unphysical
sheet and it transforms from the pole of the function ${\rm
S}_0(z)$ into its root, $E_t^{(1)*}$, corresponding to the
trimer virtual level.  The results of calculation of this
virtual level where $\lambda$ changes from 1.18 to 1.5 are
presented in Table~V.  For all the values of $\lambda$ presented
in Tables~IV and~V, the dimer possesses an only bound state. We
have found that the first excited state of the dimer appears
only at $\lambda=6.81$.

Note that in the case of finite potentials the geometric
characteristics of the domain $\Pi^{(S)}$ where the function
${\rm S}_0(z)$ can be calculated reliably, are only determined
by the value of $|\epsilon_d(\lambda)|$ (see
formula~(\ref{PiPsi}) for $\mu=+\infty$). When
$|\epsilon_d(\lambda)|$ increases, the domain $\Pi^{(S)}$ is
enlarged. It is easy to check that the energies of the excited
state level $E_t^{(1)}(\lambda)$ and of the virtual level
$E_t^{(1)*}(\lambda)$ given in Tables~IV and~V belong to the
corresponding domains $\Pi^{(S)}(\lambda)$. For $\lambda\geq1$,
this results in a weak dependence of the calculated values of
$E_t^{(1)}(\lambda)$ and $E_t^{(1)*}(\lambda)$ on the parameters
$N_\theta$, $N_\rho$ and (this is especially important) on the
parameter $\rho_{\rm max}$.

In essential, we chose the values of the cutoff hyperradius
$\rho_{\rm max}$ given in Tables~III\,--\,V from the scaling
considerations. As a matter of fact, we took the value of
$\rho_{\rm max}$ following the formula
\begin{equation}
\label{rhomax}
\rho_{\rm max}(\lambda)=\frac{C_1}{\sqrt{|\epsilon_d(\lambda)|}},
\end{equation}
where the ``constant"
$C_1=\reduction{(\sqrt{|\epsilon_d|}\,\rho_{\rm
max})}{\lambda=1}$ corresponds to an appropriate choice of
$\rho_{\rm max}$ at $\lambda=1$.  It has been established
in~\cite{KMS-PRA,MSK-CPL,KMS-JPB} that such a choice is ensured
if $\reduction{\rho_{\rm max}}{\lambda=1}=$\,400\,---\,600\,\AA.
In determining the values of $\rho_{\rm max}(\lambda)$,
indicated in Table~III, we followed the formula~(\ref{rhomax})
literally. As the ``constant" $C_1$, we took its value
corresponding to the base value of $\reduction{\rho_{\rm
max}}{\lambda=1}=$600\,\AA.  The values of $\rho_{\rm
max}(\lambda)$ presented in Tables~IV and V correspond to the
choice of $\reduction{\rho_{\rm max}}{\lambda=1}$ in the
interval within 400 and 800\,{\AA}. All the results presented in
Tables~III\,--\,V have been obtained with the grids parameters
$N_\theta=N_\rho=600$.


\bigskip
\acknowledgements

The authors are grateful to Prof.~V.\,B.\,Belyaev and Prof.~H.\,Toki
for help and assistance in calculations at the supercomputer of
the Research Center for Nuclear Physics of the Osaka University,
Japan. Also, the autors thank Mrs.~T.\,Dumbrais for her help in
translation of the text into English. One of the authors
(A.\,K.\,M.) is much indebted to Prof.~W.\,Sandhas for his
hospitality at the Universit\"at Bonn. The support of this work
by the Deutsche Forschungsgemeinschaft and Russian Foundation
for Basic Research is gratefully acknowledged.



\newpage

\begin{table}
\label{Tab-HFD-B}
\caption{
The parameters for the HFD-B $^4$He$-$$^4$He potential.}
\begin{tabular}{|r|l|}
   $\varepsilon$ (K)       &  10.948     \\
   $ r_m $ (\AA)           &  2.963      \\
   $A$                     &  184431.01  \\
   $\alpha$                & 10.43329537 \\
   $\beta$                 & $-2.27965105$ \\
   $C_6$                   & 1.36745214  \\
   $C_8$                   & 0.42123807  \\
   $C_{10}$                & 0.17473318  \\
   $D$                     & 1.4826\\
\end{tabular}
\end{table}

\begin{table}
\label{tableTrimerConvergence}
\caption
{Dependence of the difference $\epsilon_d-E_t^{(1)}$ (mK)
between the dimer energy $\epsilon_d$ and the trimer excited
state energy $E_t^{(1)}$ on the grid parameters. The values of
$\rho_{\rm max}$ are in \AA.}
\begin{tabular}{|l|c|c|c|}
\phantom{AAAAAAA}
$N_\theta,N_\rho$ $(N_\theta=N_\rho)$ & 600 & 800 & 1000 \\
$\rho_{\rm max}$ &&&\\
\hline
 400 & 0.7752 & 0.7661 & 0.7625  \\
 600 & 0.7809 & 0.7695 & 0.7649  \\
 800 & 0.7852 & 0.7723 & 0.7669  \\
\end{tabular}
\end{table}

\begin{table}
\label{tableTrimerVirtEsc}
\caption
{The dimer binding energy
$\epsilon_d$ and the differences
$\epsilon_d-E_t^{(1)}$, $\epsilon_d-E_t^{(2)}$
$\epsilon_d-E_t^{(2)*}$ and $\epsilon_d-E_t^{(2)**}$ (all in mK)
between this energy and the trimer exited-state energies
$E_t^{(1)}$, $E_t^{(2)}$ and the  virtual-state energies
$E_t^{(2)*}$, $E_t^{(2)**}$  depending on factor
$\lambda$.}
\begin{tabular}{|c|c|c|c|c|c|c|}
 $\lambda$   & $\epsilon_d$    & $\epsilon_d-E_t^{(1)}$ &
  $\epsilon_d-E_t^{(2)*}$ & $\epsilon_d-E_t^{(2)**}$ &
  $\epsilon_d-E_t^{(2)}$ & $\rho_{\rm max}$ (\AA) \\
\hline
 0.995 & $-1.160$ &  0.710       & --           & --       & -- & 723\\
 0.990 & $-0.732$ &  0.622       & --           & --       & -- & 910\\
0.9875 & $-0.555$ &  0.573       & 0.473        & 0.222    & -- & 1046\\
 0.985 & $-0.402$ &  0.518       & 0.4925       & 0.097    & -- & 1228\\
 0.980 & $-0.170$ & 0.39616      & 0.39562      & 0.009435 & -- & 1890\\
 0.975 & $-0.036$ & 0.2593674545 & 0.2593674502 & --  & 0.00156 & 4099\\
\end{tabular}
\end{table}


\begin{table}
\label{tableTrimerExcited}
\caption
{Dependence of the dimer energy
$\epsilon_d$ and the difference
$\epsilon_d-E_t^{(1)}$ between this energy and the trimer
exited-state energy  $E_t^{(1)}$ on the factor
$\lambda$.}
\begin{tabular}{|c|c|c|c|}
 $\lambda$ & $\epsilon_d$ (mK) & $\epsilon_d-E_t^{(1)}$ (mK)
& $\rho_{\rm max}$ (\AA) \\
\hline
  1.05       & $-12.244$  &   0.873   & 300  \\
  1.10       & $-32.222$  &   0.450   & 200  \\
  1.15       & $-61.280$  &   0.078   & 150  \\
  1.16       & $-68.150$  &   0.028   & 120  \\
  1.17       & $-75.367$  &   0.006   & 120  \\
\end{tabular}
\end{table}

\begin{table}
\label{tableTrimerVirtual}
\caption
{Dependence of the dimer energy
$\epsilon_d$  and the difference
$\epsilon_d-E_t^{(1)*}$ between this energy and the trimer virtual-state
energy $E_t^{(1)*}$ on the factor
$\lambda$.}

\begin{tabular}{|c|c|c|c|}
 $\lambda$ & $\epsilon_d$ (mK) & $\epsilon_d-E_t^{(1)*}$ (mK) &
$\rho_{\rm max}$ (\AA) \\
\hline
  1.18       & $-82.927$  &  0.001    & 110  \\
  1.19       & $-90.829$  &  0.016    & 110  \\
  1.20       & $-99.068$  &  0.057    & 100  \\
  1.25       & $-145.240$ &  0.588    &  85  \\
  1.30       & $-199.457$ &  1.831    &  70  \\
  1.35       & $-261.393$ &  3.602    &  70  \\
  1.40       & $-330.737$ &  6.104    &  55  \\
  1.50       & $-490.479$ &  12.276   &  50  \\
\end{tabular}
\end{table}

\newpage


%
\begin{figure}
\centering
\epsfig{file=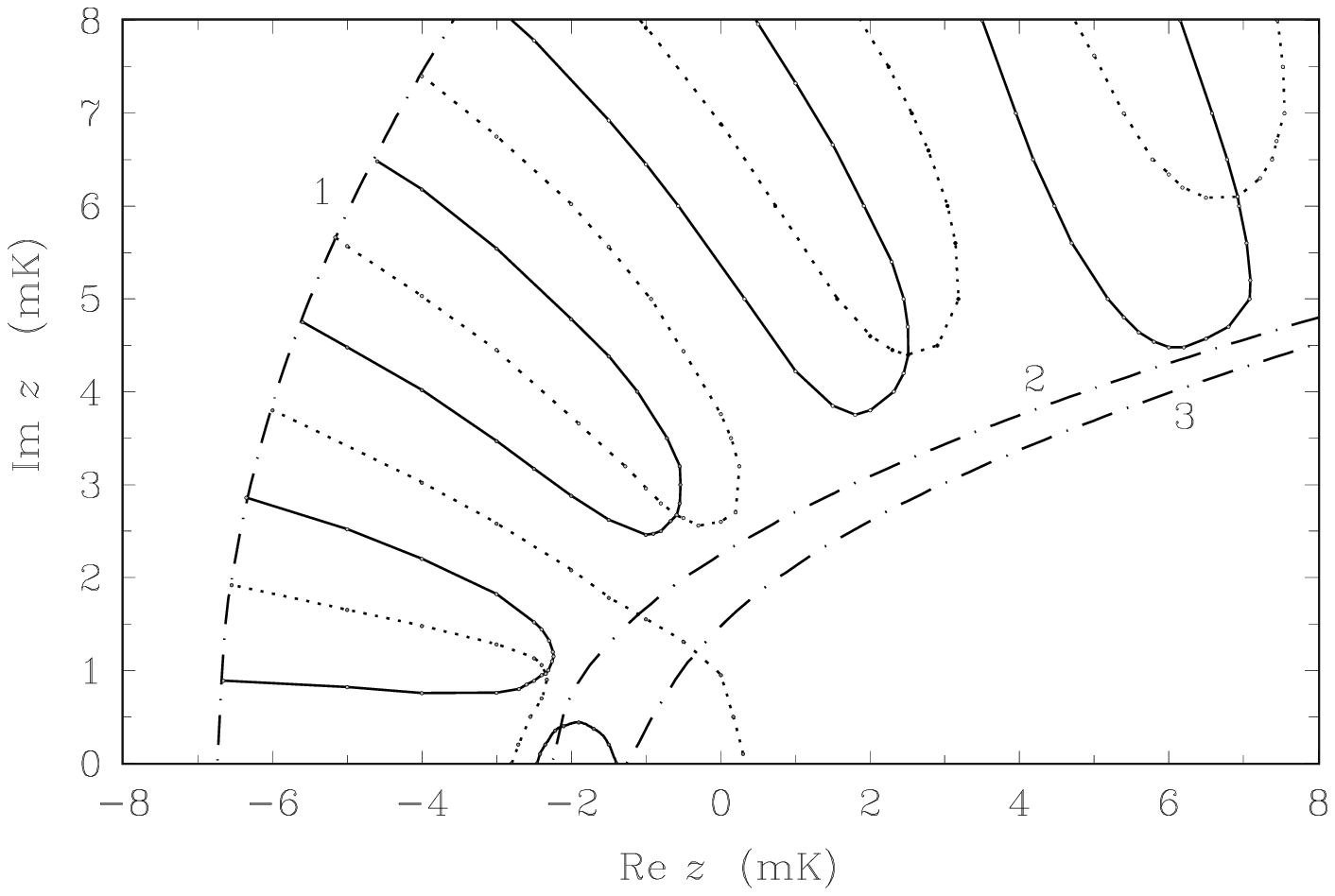,height=12cm}
\caption{Root locus curves of the real and imaginary parts of
the scattering matrix ${\rm S}_0(z)$. The solid lines correspond
to $\Real {\rm S}_0(z)=0$, while the tiny dashed lines, to $\Img
{\rm S}_0(z)=0$.  The Numbers 1, 2, 3 denote the boundaries of
the domains $\Pi^{(\Psi)}$, $\Pi^{(S)}$ and $\Pi^{(A)}$,
respectively. Complex roots of the function ${\rm S}_0(z)$ are
represented by the crossing points of the curves $\Real {\rm
S}_0(z)=0$ and $\Img {\rm S}_0(z)=0$ and are located at
\mbox{$(-2.34+{\rm i}\,0.96)$}\,mK, \mbox{$(-0.59+{\rm
i}\,2.67)$}\,mK, \mbox{$(2.51+{\rm i}\,4.34)$}\,mK and
\mbox{$(6.92+{\rm i}\,6.10)$}\,mK.}
\label{RootLines}
\end{figure}

%
\begin{figure}
\centering
\epsfig{file=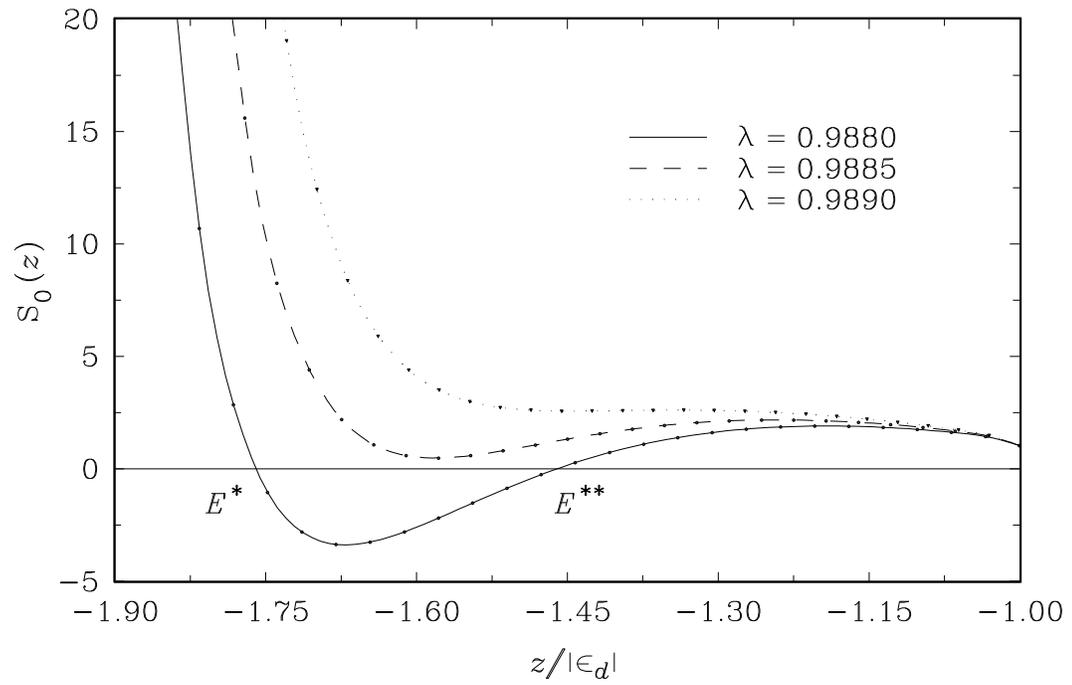,height=12cm}
\caption{Graphs of the function ${\rm S}_0(z)$ at real
$z\leq\epsilon_d$  for three values of $\lambda<1$ .
The notations used: $E^{*}=E_{t}^{(2)*}/|\epsilon_d|$,
$E^{**}=E_{t}^{(2)**}/|\epsilon_d|$.}
\label{S0Virt}
\end{figure}
\end{document}